\documentclass[aps,twocolumn,amsmath,amssymb,pre]{revtex4}

\usepackage[dvips]{graphicx}

\newcommand{\be}{\begin{equation}}
\newcommand{\ee}{\end{equation}}
\begin{document}

\title{Reply to the comment of T.Gilbert and D.P.Sanders on
``Capturing correlations in chaotic diffusion by approximation
methods''}

\author{Rainer Klages}
\email{g.knight@qmul.ac.uk}
\affiliation{School of Mathematical Sciences, Queen Mary University of London,
Mile End Road, London E1 4NS, UK}
\author{Georgie Knight}
\email{r.klages@qmul.ac.uk}
\affiliation{School of Mathematical Sciences, Queen Mary University of London,
Mile End Road, London E1 4NS, UK}

\begin{abstract}

This is a reply to the comment by Gilbert and Sanders [arXiv:1111.6271
(2011)]. We point out that their comment is a follow-up of a previous
discussion which we briefly summarize before we refute their new
criticism.

\end{abstract}

\maketitle

\noindent In their comment \cite{GilSan11} Gilbert and Sanders argue
that our application of persistent random walk theory to maps
\cite{KnKl11b}, in response to their reference \cite{GilSan09}, is
`inaccurate', due to a `misinterpretation' of simple random walk
theory, and `based on a confusion between two different time
scales'. Their comment is a follow-up of a discussion that has started
with \cite{KlKo02} authored by one of us. This work was already
criticized by Gilbert and Sanders in \cite{GilSan09} on which we
responded with \cite{KnKl11b}. We welcome the opportunity to briefly
summarize this discussion in order to put the present comment by
Gilbert and Sanders into context before we refute their new criticism.

In \cite{KlKo02} the Taylor-Green-Kubo formula, which expresses
the diffusion coefficient in terms of the velocity autocorrelation
function, has been worked out for periodic particle billiards by
mapping the time-continuous dynamics onto a time-discrete correlated
random walk on a periodic lattice. This approach expresses the
diffusion coefficient in the form of a series expansion that converges
exactly. Calculating the $k$-th order terms of this expansion yields a
systematic approximation procedure for the diffusion
coefficient. Non-trivial information is provided by this method in
form of the parameter dependence of the convergence, as was
demonstrated in \cite{KlKo02} for one-dimensional maps and for
billiards both numerically and analytically. This conceptually trivial
method was successfully applied to a number of different dynamical
systems in order to physically understand the irregular parameter
dependence of diffusion coefficients, see \cite{Kla06} and
further references therein.

In a first round of criticism Gilbert and Sanders argued that this
method is mathematically wrong and physically meaningless
\cite{GilSan09}. They claimed that in order to obtain `accurate
results' for capturing correlations one must use persistent random
walk theory.  A substantial part of this theory was originally
developed for correlated random walks on lattices (see, e.g.,
\cite{OBFG80,HK87} and further references therein). By employing
the Taylor-Green-Kubo formalism of \cite{KlKo02}, Gilbert and
Sanders adapted this theory to billiards and demonstrated its
application by numerical results \cite{GilSan09}.

This criticism of Gilbert and Sanders was refuted in \cite{KnKl11b} by
analyzing diffusion in a simple one-dimensional chaotic map. This
model had the advantage that in contrast to billiards it is amenable
to rigorous analysis. Here both the Taylor-Green-Kubo approximation
method of \cite{KlKo02} and a suitable adaption of persistent random
walk theory to maps (plus another new, third method) were applied and
compared with each other. Our results re-confirmed that the criticism
by Gilbert and Sanders was unfounded, as could already have been
inferred from \cite{KlKo02,Kla06} and further references therein. The
present comment by Gilbert and Sanders on \cite{KnKl11b} does not
contain any further evidence that this approach is incorrect. We are
thus pleased to conclude that Gilbert and Sanders have accepted the
invalidity of their previous criticism published in \cite{GilSan09}.

Instead, with their comment \cite{GilSan11} Gilbert and Sanders launch
a second round of criticism on a new aspect that is still related to
the Taylor-Green-Kubo formalism but different from the previously
criticized approximation method.  Their argument is summarized as
follows: Transposing persistent random walk theory to billiards on the
basis of \cite{KlKo02} led to a correlated random walk dynamics
which did not incorporate a probability that particles on the random
walk lattice remain at a site \cite{GilSan09}. The new criticism of
Gilbert and Sanders is that, by working out persistent random walk
theory for time-discrete maps defined on a periodic lattice, we
included a non-zero probability that particles remain on a site. In
view of their own adaption of persistent random walk theory to
billiards \cite{GilSan09}, which motivated \cite{KnKl11b}, they
argued that we have `confused' two different time scales in our
analysis, namely `the average time spent by a walker at any given site
before it moves on to a neighboring one' with the `unit time scale of
the underlying process'.

Our reply to their criticism is as follows: We first re-emphasize that
persistent random walk theory has to a large extent originally been
developed for time-discrete stochastic random walks on periodic
lattices \cite{HK87,OBFG80}. The one-dimensional map studied in
\cite{KnKl11b} is a simple deterministic realization of such a random
walk. Gilbert and Sanders's reference \cite{GilSan09} suggests a
suitable adaption of persistent random walk theory to billiards. In
their comment they now insist that their billiard version must be
re-transposed to maps by ruling out any non-zero probability that a
walker remains on a site. This constraint is at variance with original
persistent random walk theory, which explicitly does incorporate such
probabilities \cite{OBFG80}, as motivated by applying this theory to
experimental data \cite{Sal90}. Furthermore, in a follow-up article
\cite{GNS11} Gilbert et al.\ reproduce results that are contained as a
special case in a more general formula derived in \cite{OBFG80}, which
incorporates a non-zero probability that a walker remains on a site,
compare Eqs.(1.3) and (5.4) in \cite{OBFG80} with (2.17) in
\cite{GNS11}. Our results for maps \cite{KnKl11b} are consequently
neither `inaccurate', nor is there any `flaw' in view of standard
persistent random walk theory. Secondly, there is no `misunderstanding
of the Machta-Zwanzig approximation', i.e., simple uncorrelated random
walk theory for billiards \cite{Kla06}: As was shown in
\cite{KnKl11b}, this approximation is exactly reproduced by our method
for our model, cf.\ Eq.(26). In summary, there is neither any
contradiction between persistent random walk theory and how we apply
it to maps, nor is there any error in our theory, in contrast to what
Gilbert and Sanders suggest in their comment.

We finally address the criticism of Gilbert and Sanders of having
`confused' two different time scales: A key observation in
\cite{KlKo02} was that the natural time scale in order to map
time-continuous diffusion in billiards onto a time-discrete random
walk on a periodic lattice is the average escape time $\tau$ of a
particle to move from one fundamental cell to another. For
time-discrete diffusion on periodic lattices, as for time-discrete
maps, the natural time scale is the unit time, possibly multiplied by
some factor. It is thus convenient to choose these natural time scales
for working out persistent random walk theory in these different
systems, and there is no inaccuracy in doing so. Another important
point is that in \cite{KnKl11b} we compare three different
approximation schemes for diffusion with each other. In order to
provide a fair comparison between these different methods, one needs
to set them up by using the very same time scale. Persistent random
walk theory and the Taylor-Green-Kubo approximation method are based
on the very same Taylor-Green-Kubo formula, which guarantees that this
condition is fulfilled by default. In \cite{KnKl11b} we have
conveniently chosen the unit time as a time scale therein. Hence,
instead of any confusion, there was deliberately a clear choice of a
unique time scale. The comment by Gilbert and Sanders shows that by
using the average escape time $\tau$ for maps instead of unit time,
the textbook uncorrelated random walk result matches to the lowest
order of persistent random walk theory with zero probability of
trapping. We encourage Gilbert and Sanders to go beyond this lowest
order by re-calculating our {\em higher-order corrections} for {\em
all three approximation methods} \cite{KnKl11b} using a non-integer
time scale $\tau$ for time-discrete maps.

We conclude our reply by pointing out that, as Gilbert and Sanders
have already misread \cite{GilSan09} the Taylor-Green-Kubo
approximation scheme developed in \cite{KlKo02}, they have also
misread \cite{GilSan11} reference \cite{KnKl11b}: Nowhere in our
article do we suggest that there is `a serious limitation of [Gilbert
and Sanders's] formalism' \cite{GilSan09} due to zero velocity
states. Such states merely prevented us from obtaining an analytical
solution for the two-step approximation of our model, cf.\ the very
last sentence in the Appendix of \cite{KnKl11b}, a problem that was
resolved by reverting to numerical methods. The comment by Gilbert and
Sanders is thus motivated by an inadequate interpretation of our
article. A correct account is that we have compared three different
approximation methods in it by applying them on the same footing,
i.e., by using the very same fundamental time scale, to a given
map. For this model we conclude that all these schemes have their own
virtues and deficiencies. We thus repeat again, as we concluded
already in \cite{KnKl11b}, that we find `the quest for a unique way to
approximate the diffusion coefficient of a dynamical system', as
repeatedly suggested by Gilbert and Sanders by pointing towards their
own adaption of persistent random walk theory to billiards,
`unnecessarily restrictive'.

%\bibliography{summ15}

\end{document}